**RESEARCH ARTICLE**             **OPEN ACCESS**

# Identifying Structures in Social Conversations in NSCLC Patients through the Semi-Automatic extraction of Topical Taxonomies

Giancarlo Crocetti*, Amir A. Delay**, Fatemeh Seyedmendhi***
*(Department of Computer Science, St. John's University)
**(Department of Computer Science, St. John's University)
***(Department of Computer Science, St. John's University)

**ABSTRACT**
The exploration of social conversations for addressing patient's needs is an important analytical task in which many scholarly publications are contributing to fill the knowledge gap in this area. The main difficulty remains the inability to turn such contributions into pragmatic processes the pharmaceutical industry can leverage in order to generate insight from social media data, which can be considered as one of the most challenging source of information available today due to its sheer volume and noise. This study is based on the work by Scott Spangler and Jeffrey Kreulen and applies it to identify structure in social media through the extraction of a topical taxonomy able to capture the latent knowledge in social conversations in health-related sites. The mechanism for automatically identifying and generating a taxonomy from social conversations is developed and pressured tested using public data from media sites focused on the needs of cancer patients and their families. Moreover, a novel method for generating the category's label and the determination of an optimal number of categories is presented which extends Scott and Jeffrey's research in a meaningful way. We assume the reader is familiar with taxonomies, what they are and how they are used.
***Keywords*** – Taxonomy, Social Media, Text Mining, Clustering, Knowledge Management

## I. Introduction

The availability of taxonomies varies dramatically from domain to domain. In life sciences, for example, we have an abundance of well-curated and up-to-date taxonomies (e.g. MeSH, MeDRA, Entrez Gene, etc.), from medical to genetic to disease terminologies. However, the situation is very different in other domains. With the exception of the financial area, it is very hard to come by domain-specific taxonomies, and when available, they are often proprietary with a hefty price tag. The problem is further exacerbated when we are interested in ascertaining the context around social conversation through the use of topical taxonomies which are domain specific and practically inexistent on the market for purchasing

In this situation there are not many options. You might develop your own taxonomy - after all there is a well-defined framework that can guide you through the process -, adopt an automatic strategy, or use a hybrid approach with the initial generation of terms using data mining techniques which are further reviewed by a professional taxonomist. The latter is an approach taken in this case study, in which the term taxonomy is used in its broader sense to reference any means of organizing concepts of knowledge.

The results presented in this study are based on the work done by Scott Spangler and Jeffrey Kreulen [1] and extended by deriving two methods for:

1. Identifying a possible solution for the number of categories in the extracted taxonomy.
2. labeling each category using descriptors within the cluster

This work is focused on the extraction of a taxonomy on social data, but it can easily be extended to any other domain.

The development of a new taxonomy requires a significant effort thatnot only ends with the result of the initial taxonomy, but will also require continual maintenance and governance. Several authors such as Patrick Lambe [2] and Heather Hedden [3]outline a process for the development of a taxonomy. Other processes exist in the industry, and together they define a multi-step procedure for developing an in-house taxonomy which usually requires:

1. The definition of a business case
2. The engagement of stakeholders
3. Having in place a strong communication plan
4. The identification of the design
5. The definition of a governance model.

When compared to other designs, the automatic extraction of a draft taxonomy might appeal for several reasons:

1. Domain experts might not be readily available in your company; therefore, data mining approaches might represent a viable solution.
2. You might have time constraints on your project that does not allow the running of workshops or interviews.
3. Budget constraints.

## II. Literature Review

Social media analysis is an interest for researchers because sites related to healthcare societies are in need to compare previous medical records. This can be used by practitioners for getting information related to patient care and overall productivity. Other areas that maybe of interest are promotional information, addressing confusing terminologies, increasing communication between patients, families, and having more of a variety of people who participate in health research and clinical trials. Therefore, we can use this to raise the bidirectional stream of communication between the doctors and their patients. This can be used to update the physician's use of social media as a proficient way to share new medical information within the medical community and also develop healthcare quality[4].

In order to discover important topics in forums related to health information and patients' information needs for healthcare knowledge, we need to analyze content identification. We can use methods, such as surveys based on questionnaires and statistical analysis. In previous studies, analysis of information that are shared in medical support blogs were based on the number of people who used it and the frequency of postings. Using this information from the surveys they were able to evaluate different user groups from the data[5].

However, there were some problems in these statistical methods; the sample populations that researchers chose were too narrow and bias, so it affected the accuracy of their model. Another problem that arose was the fast development of these health forums and websites; they needed another method that can handle these lager amounts of data and be able to process at optimal speeds[5].

Previous studies demonstrated that the most frequent themes with which patients are concerned are prevention, diagnosis, support, treatments and the long-term side effects of those treatments. However, statistical content analysis is based on human defined content, which requires significant amounts of effort. This can be a time consuming process which is often error-prone and time consuming. In fact, when using traditional statistical methods, we have difficulty discovering relationships within data. Thus these methods are unpractical due to time and storage constraints[5].

To solve this, clustering methods seems to provide a good alternative to purely statistical methods. Clustering is primarily used to find unique relationships and patterns within large datasets that previously have no organization. This technique has been used in complicated task such as, pattern recognition, image analysis, and facility location. They are able to use clustering in order to partition the data into homogenous clusters, which are based on the content of the data, and give us an un-biased approach to looking at our content[6].

In fact, cluster analysis refers to an area of multivariate statistics that involves the grouping of objects based on some measure of proximity defined amongst those objects. Unlike discriminant analysis, which assumes that the group memberships are known, cluster analysis generates group memberships based on the proximities of data. Cluster analysis also differs markedly from principal component analysis and factor analysis. Whereas principal component analysis and factor analysis typically focus on reducing dimensionality by establishing linear combinations of variables; cluster analysis centers on classification of the objects based on their proximity with respect to variable measurements. Lu and colleagues [5] proposed a method based on clustering to explore health-related discussions in online health communities automatically instead of using the statistical approaches employed in previous studies. By integrating medical domain-specific knowledge, they have constructed a medical topic analysis model. The application of clustering for the extraction of taxonomies was successful also in other areas such as the automotive industry[7]. Ringsquandl presented a novel approach to semi-automatically learn concept Hierarchies from natural language requirements of the automotive industry. They extract taxonomies by using clustering techniques in combination with general thesauri. Evaluation shows that this taxonomy extraction approach outperforms common hierarchical clustering techniques[7]. Han, proposed a hierarchical clustering algorithm based on an asymmetric distance metric to explore hierarchical folksonomy for social media[8].

## III. Methodology

Without considering types of taxonomies you need to focus on relations and the fact they can contain different structures and different kinds of relationships between terms.The basic building blocks in any taxonomy remain the set of terms; therefore, the first step is to identify such set of terms.

It shows from the literature review that clustering algorithms have a proven record on the identification of descriptive terms; it is not a coincidence that this was the method of choice for Spangler and Kreulen [1]. Consequently, we used the same k-Means algorithm as a starting point so to create a baseline that could be further improved with the use of other clustering approaches which are outside the scope of this study.

The decision to use the k-Means was based on the ability of this algorithm to create centroids which can be considered as a summary of all conversations contained in a cluster, generated as the arithmetic average of its elements. These centroids, together with a measure of similarity, provide a good platform to identify key terminology in the grouping of similar conversations. Despite its simplicity, we have to be aware of some of its major drawback related to the identification of sub-optimal solutions. Because, at this point, we only interested in testing a candidate process, we are not particularly concerned to its optimization tasks and the use of an algorithm that is very transparent in its execution and result generation it is preferable to other, more opaque, methods.

The goal of this process is to create a possible grouping of terms, organized in a one-level hierarchy, from a topic-specific collection of social media posts in order to generate a draft taxonomy able to capture the important knowledge expressed in these social conversations. Each level is represented by a single cluster and the associated descriptors are extracted using the centroid information generated for each cluster.

We are well aware of the current impossibility to generate taxonomies using completely automated means and the support of an expert taxonomist is of paramount importance. Consequently, the entire process has been developed so that the time needed to review the resulting "draft" taxonomy by a domain expert, wound not take more than one hour. This can only be achieved but setting constraints on the number of categories and descriptors included in the taxonomy, yet, it is not easy to determine such thresholds A good number for categories is 30 which allows the analyst to visualize them at once in tabular form. With this first constraint in place we can set the maximum number of descriptors to 2,000. This number is derived considering 30 categories containing no more than 29 unique descriptors for a total of 870 features. To be on the safe side, Spangler and Kreulen doubled this number to 1740 and round it up to a maximum of 2,000 features which is a very reasonable number.

A taxonomy of this size can be easily reviewed by an expert taxonomist within one hour. We will also see that, in particular, the constraints on 30 categories will allow the optimization on the number of clusters for a given solution.

### 3.1 Source of data
In this study we crawled conversations from social sites that do not require login and with open access to their forums. In particular we collected data from: www.cancerforums.net, www.lunglovelink.org, and www.lungevity.org. Among the forums available for crawling we collected social posts related to Small Cell Lung Cancer using an in-house crawler developed using Python, based on the open source Beautiful Soup, and stored the social data in the form of XML documents with a canonical structure independent of any particular site. Once the crawling was completed the XML data was converted into a comma delimited file for further processing.

The final dataset contained a total of posts with the breakdown shown in Figure 1.

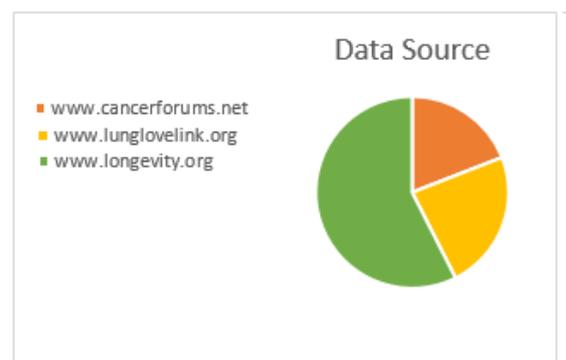

Figure 1 - Proportion of posts for each data source

### 3.2 Extracting and preparing the data
The data mining tool used in this experiment is RapidMiner with the addition of the TextMining add-on that is freely available and downloadable within the tool itself.
The process of text preprocessing in social posts is described in Figure 2.

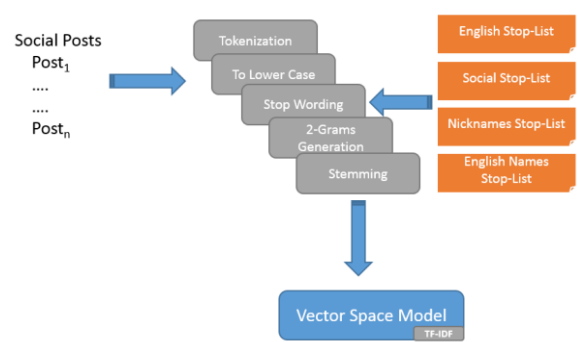

Figure 2 - Preprocessing of Social Posts

As in any text mining pre-processing task, we began the processing by transforming the social conversations into the well-known Vector Space Model by using the "Process Document" operator in RapidMiner using TF-IDF measures of frequency.

We first utilized tokenization, which is a text processing methods that separates text into a sequence of tokens (single words) using the spaces between words and removing any punctuation character. Next we transformed the tokens to contain all lowercase letters so that word frequency can be accounted for.

Social conversations represent very noisy data, with the use of many embellishing terms that are weakly related to the topic of the conversation. Therefore, special attention must be paid to the removal of such noise. To this end, we applied four stop-wording dictionaries that are specific to remove the high frequency English words such as words that are commonly used in cancer discussions, the list of usernames, and the first names of people in the discussion.

2-grams are therefore generated to capture pair of terms that have special meaning when used together. For small documents like social interactions it does not make much sense to go beyond 2-grams. Only the 2-grams with a minimum document frequency of 3 (they must appear in at least three documents) were kept, all the other were filtered out. The resulting vocabulary contained terms like "lung_cancer", "cat_scan", or "anti_nausea" which are all examples of a meaningful terminology in the context of this experiment.

The pre-processing step generated what is commonly called the Vector Space Model (VSM) which is a matrix containing as rows the documents representing the social posts and as columns the terms extracted during the process as shown in Figure 3.

Each cell in the matrix represents the TF-IDF measure of frequency associated to that term. A discussion on TF-IDF weighting is outside the scope of this book, but this should suffice to say that this score is higher for important terms, that is, terms that appear on a restricted number of documents and possibly providing context to the social conversations in which they appear.

Figure 3 - VSM of Social Conversations

### 3.3 k-Means clustering

Organizing the data into meaningful categories is a fundamental task in the generation of a taxonomy and many approaches exists today. Clustering analysis is the formal study of methods for grouping objects based on similar characteristics so to explore their intrinsic structure. However, clustering methods do not generate labels and this shortcoming was addressed in this study.

The k-Means clustering algorithm was proposed over 50 years ago and still widely used today, even though many other algorithms have been proposed since its introduction [9]. Because our study is based on the work of Spangler and Kreulen [1], we will use the same approach that utilizes the k-Means algorithm.

### 3.4 Determining the number of categories

The application of the k-Means clustering algorithm requires the user to specify the initial number *k* of categories which, probably, is not something that is known a priori. In this particular case the help of a subject matter expert might not provide a useful insight in inferring the number of taxonomical categories contained in the social conversations under analysis.

Luckily, we can put the constraint of generating no more than 30 categories to good use. Before doing that we need to define a score that is able to measure the quality of a clustering solution so to be able to determine when a solution is better than another.

In general, we can characterize a "good solution" as a set of clusters that are far apart from each other with points within the same cluster that are as close as possible. This characterization allows us to define the following quantities [10]:

**Within Cluster Variation (WCV)**: measures how close or similar are elements within the clusters. We want this quantity to be as small as possible.

**Between Cluster Variation (BCV)**: measures how far are the clusters from each other. We want this quantity to be as large as possible.

The WCV and BCV can be represented graphically as in Figure 4.

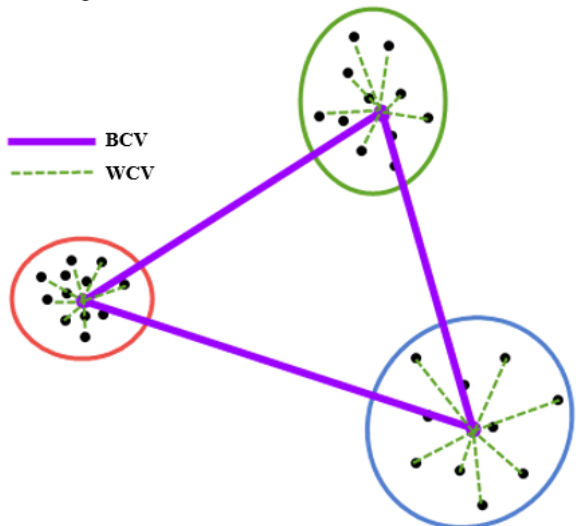

Figure 4 - WCV and BCV variation in a clustering solution

Instead of analyzing these two measures independently we can consider the quality ratio:
(1) $Q_k = BCV_k / WCV_k$ where WCV is:
(2) $\sum_{i=1}^{k} \sum_{p \in C_i} d(p, m_i)^2$ the sum of the square distances of each point in the cluster $C_i$ from the respective centroid $m_i$. We can consider this as a good approximation for the overall WCV.

Similarly, we can approximate BCV as:
(3) $\sum_{i=1}^{k} \sum_{j=1}^{k} d(m_i, m_j)^2 / 2$ with $m_i$ the centroid for cluster $C_i$. The distant function used for the calculation of WCV and BCV is the cosine similarity which works particularly well with text [11].
In general we want to have a small WCV and a large BCV, consequently we want to maximize the quality ratio $Q_k$.

From the constraint on the number of categories we know that our optimal value of $k$ is between 2 and 30. We can, therefore, run the $k$-Means algorithm for each of the possible value of $k$ between 2 and 30 and calculate the associate $q_i$ value, with $i=2, ..., 30$. Our optimal taxonomy, under such constraints, will be the one having $k$ categories so that $k: q_k = Max(q_2, ..., q_{30})$.

In Figure 5 we see the values of $Q_k$ for different values of $k$. The highest value of $q$ is associated with the solution comprised of $k=12$ clusters.

| K | WCV | BCV | q =BCV/WCV |
|---|---|---|---|
| 2 | 0.033885 | 0.000496 | 68.26 |
| 3 | 0.085983 | 0.003719446 | 23.10 |
| 4 | 0.15541 | 0.001824 | 85.20 |
| 5 | 0.157201 | 0.006031 | 26.10 |
| 6 | 0.237921 | 0.005559 | 42.80 |
| 7 | 0.299438 | 0.005442 | 55.00 |
| 8 | 0.42604 | 0.008192 | 52.00 |
| 9 | 329164.7 | 917960.3 | 0.36 |
| 10 | 0.736095 | 0.010893 | 67.60 |
| 11 | 0.818156 | 0.009247 | 88.50 |
| 12 | 0.906767 | 0.009742 | 93.10 |
| 13 | 0.753285 | 0.010636 | 70.80 |
| 14 | 0.846446 | 0.010701 | 0.08 |
| 15 | 0.998 | 0.010897 | 0.08 |

Figure 5 - Values of F statistics at different value of $k$

This is an interesting finding: the same solution was identified by a domain experts when considering the grouping for the different values of $k$. It would be interesting to support this finding with more experiments.

**3.5 Extraction of the taxonomy**
We have to identify the clustering solution for which we are getting the best separation between conversations belonging to different clusters while trying to increase similarity in conversations belonging to the same group.

It is possible, now, to generate the category label and assign the descriptors within the category. To this end we will use the centroid vector in each of the 12 clusters as in Figure 6, where we are showing the centroids for first 5 clusters and the associated TF-IDF weighting.

| Attribute | cluster_1 | cluster_2 | cluster_3 | cluster_4 | cluster_5 | cl |
|---|---|---|---|---|---|---|
| support | 0.121 | 0.004 | 0.006 | 0.016 | 0.019 | 0.0 |
| share | 0.085 | 0 | 0 | 0.004 | 0.005 | 0 |
| word | 0.064 | 0.012 | 0.006 | 0.001 | 0.006 | 0.0 |
| wonder | 0.060 | 0.030 | 0.008 | 0.004 | 0.003 | 0.0 |
| love | 0.046 | 0 | 0 | 0.004 | 0 | 0 |
| encourag | 0.044 | 0 | 0 | 0.003 | 0 | 0.0 |
| journey | 0.037 | 0.019 | 0.002 | 0.015 | 0.011 | 0.0 |
| advic | 0.037 | 0 | 0.007 | 0.009 | 0.005 | 0 |
| famili | 0.033 | 0.591 | 0.005 | 0.008 | 0.019 | 0.0 |
| site | 0.033 | 0 | 0 | 0.005 | 0 | 0 |
| comfort | 0.032 | 0 | 0.002 | 0.017 | 0 | 0.0 |
| live | 0.028 | 0.005 | 0.010 | 0.003 | 0.008 | 0.0 |
| care | 0.027 | 0.033 | 0.018 | 0.007 | 0.015 | 0.0 |
| friend | 0.025 | 0 | 0 | 0.002 | 0.007 | 0 |
| help | 0.024 | 0 | 0.016 | 0.012 | 0.007 | 0.0 |
| learn | 0.023 | 0 | 0.002 | 0.001 | 0.007 | 0 |
| life | 0.023 | 0.004 | 0.003 | 0.004 | 0 | 0.0 |

Figure 6 - Example of centroid vectors

3.5.1 Extraction category labels

The label for each category is created through the concatenation of terms within each centroid, which poses the following issues:

1. Which terms should we pick?
2. How many terms should we consider?

To address these issues we consider the centroid vector $m_i$ for each of the cluster $c_i$ and sort its terms by the TF-IDF values in descending order as shown in Figure 7.

| Attribute | cluster_3 |
|---|---|
| nausea | 0.178 |
| chemo | 0.163 |
| anti_nausea | 0.099 |
| anti | 0.097 |
| med | 0.087 |
| nausea_m | 0.071 |
| drug | 0.056 |
| feel | 0.045 |
| treatment | 0.044 |

Figure 7 - Terms sorted by their TF-IDF weights

The procedure to create the category label is then the following:

1. Include the first term $t_1$ in the label
2. Consider the ratio $\frac{tf-idf_{i+1}}{tf-idf_i}$ of the TF-IDF weights of two subsequent terms $t_i$ and $t_{i+1}$
3. Add the term $t_i$ to the label if $\frac{tf-idf_{i+1}}{tf-idf_i} > 0.5$
4. Repeat step 2 and 3 until it is possible to add items without discontinuity

The application of this procedure to the cluster in Figure 7, generates the label "Nausea & Chemo & Anti_Nausea", and in Table 1, we reported the labels generated for the entire cluster solution.

Table 1 – Labels generated by the algorithm

| Cluster Number | Cluster Labels (Category Label) |
|---|---|
| 0 | news & famili & happi |
| 1 | stori & share & inspir |
| 2 | support & peopl & lung |
| 3 | nausea & chemo & anti_nausea |
| 4 | Alimta |
| 5 | scan & nodul & ct_scan |
| 6 | zometa |
| 7 | lung & surgeri & tumor |
| 8 | chemo & treatment & radiat |
| 9 | thread & repli & tarceva |
| 10 | comfort & pass & heart |
| 11 | pain & feel & tri |

Notice how the algorithm was able to identify conversation topics around specific drugs like Alimta, Zometa, and Tarceva, which are used in the treatment of Non-Small Lung Cancer and might represent an important insight in the analysis of conversations around this topic, especially if we are interested in patient's perception on the treatment of NSLC.

3.5.2 Filling the category with descriptors
With the category labels in place we can now add the remaining list of terms in the centroid vectors as descriptors to further detail the conversation's topics identified previously. In considering the example we used before, related to cluster 3 (see Figure 7), we have the following category.

**Nausea & Chemo & Anti_Nausea**
Chemo
Drug
Medication
Treatment
anti-nausea
nausea medication
nausea
…
This is a well-defined concept in which patients are clearly conversing on the side effect of chemotherapy and possible solutions.
Let's look at another example as in cluster 8: "Chemo & Treatment & Radat":

**Chemo & Treatment & Radiat**
Chemo
Treatment
Radiat
Tarceva
Doctor
Trial
…

Which relates to conversation around trials for the use of Tarceva in addition to radiation therapy. The topical insight is quite strong in this case as well.

The resulting on-level categorization generated under the constraints highlighted in this methodology can be easily reviewed by a domain expert within one hour and the resulting curated taxonomy used to identify specific topics in social conversation.

## IV.     Conclusions and further work

The interesting work of Spangler and Kreulen [1] related to the analysis of social conversations has been extended in important ways through the addition of several improvements:

1. A better insight of the number of clusters *k* inputted in the k-Means clustering algorithm
2. Generation of categorical labels related to the grouping of conversations

The labels generated by the procedure illustrated in this study can provide an immediate insight on the topics discussed in social media posts without the need of reading this vast amount of text data. Moreover, the curated taxonomy outputted by this procedure can be used by annotation services able to capture important terminology in social media data sources related to the topics described in the taxonomy.

The methodology in this study was based on the same clustering approach, using k-Means, by Spangler and Kreulen and it would be interesting to have comparative study using other algorithmic models. Moreover the resulting taxonomy at the end of the process is one-level deep, which could be extended with the application of recursive methods and improved through the use of lexicon resources so to derive broader/narrower relationships and the addition of synonyms.

## V.     Aknowledgments

This research was not possible without the help and support from the following graduate students at the Seidenberg School of Computer Science and Information Systems of Pace University: Mr. Suhail Bandhari, Ms. Trishna Mitra, Ms. Manvi Mohta, and Daniel Rings. Their help in identifying the social media sites and extracting the data was invaluable. A special thanks also to Prof. Catherine Dwyer for her support in this project and for coordinating the students.

## References

[1] S. Spangler and J. Kreulen, *Mining the Talk: Unlocking the Business Value in Unstructured Information* (Boston, MA: IBM Press, 2007).
[2] P. Lambe, *Organising Knowledge: Taxonomies, Knowledge and Organisational Effectiveness* (Oxford, UK: Chandos Publishing, 2007)
[3] H. Hedden, *The Accidental Taxonomist* (Medford, NJ: Information Today, 2011)
[4] B.S. McGowan, M. Wasko, S.B. Vartabedian, S.R. Miller, D.D. Freiherr, and M. Abdolrasulnia, Understanding the Factors that Influence the Adoption and Meaningful Use of Social Media by Physicians to Share Medical Information, *Journal of Medical Internet Research, 14(5),* 2012.
[5] Y. Lu, P. Zhang, J. Liu, J. Deng, and S. Li, Health-Related Hot Topic Detection in Online Communities Using Text Clustering, *PLOS ONE, 8(2),* 2013.
[6] W. Yeh and C. Lai, Accelerated Simplified Swarm Optimization with Exploitation Search Scheme for Data Custering, *PLOS ONE,* 2015.
[7] M. Ringsquandl and M. Schraps, Taxonomy Extraction from Automotive Natural Language Requirements Using Unsupervised Learning, International,*Journal on Natural Language Computing, 8(4),* 2014.
[8] Z. Han, Q. Mo, Y, Liu, and M. Zuo, Contructing taxonomy by hierarchical clustering in online social bookmarking, *Educational and Information Technology, 3,* 2010, 47-51.
[9] A. K. Jain, Data Clustering: 50 years beyond k-Means, *Pattern Recognitions letter, 31(8),* 2010, 651-666.
[10] D. Larose and C. Larose, *Discovering Knowledge in Data* (Hoboken, NJ: Wiley, 2015)
[11] A. Huang, Similarity Measures for Text Document Clustering, *Proceeding of the New Zealand Computer Science Research Student*, 2007